# Criminal Geographical Profiling: Using FCA for Visualization and Analysis of Crime Data

Quist-Aphetsi Kester, MIEEE
Lecturer, Faculty of Informatics
Ghana Technology University College Accra, Ghana
Email: kquist-aphetsi@gtuc.edu.gh / kquist@ieee.org

*Abstract—fighting criminal activities in our modern societies required the engagement of intelligent information systems that can analyze crime data geographically and enable new concepts to be deduced from it. These information systems should be able to create visualization of such data as well as have the capability of giving new incite of information, if data is updated whilst maintaining the previously predicted patterns.*
*This paper proposed the use of Formal Concept Analysis, or Galois Lattices, a data analysis technique grounded on Lattice Theory and Propositional Calculus, for the visualization and analysis of crime data. This method considered the set of common and distinct attributes of crimes in such a way that categorization are done based on related crime types, geographical locations and the persons involved.*

## I. INTRODUCTION

To be able to fight crime effectively or understand criminal activities in the future, our information systems need to have the capability of analyzing and extracting significant knowledge from data based on predefined rules with supervised or unsupervised learning techniques. Hence there is a need for us to use effective computational and mathematical models in the domain of data mining and machine learning in building our artificial intelligence systems. These information systems should have the capability of using economic data, geographical data, demographic data, social networking data etc in analyzing and predicting behavior of modern society especially in the domain of peace and security which is one of the crucial platforms for effective development.

Modern society has been challenged with rise in criminal activities and these crime rates vary enormously from one country to another and from one region to another [1].With the documentation of criminal activities and the use of computerized systems to track crimes, computer data analysts have started helping the law enforcement officers etc to understand crime patterns [2]. These systems should be capable of gathering and interpreting intelligence so as to help control of criminal activities as well as influence effective decision making as in figure 1 below [3]. Since criminal activities have become very complex, its monitory with intelligent systems has become necessary by using with geographical components.

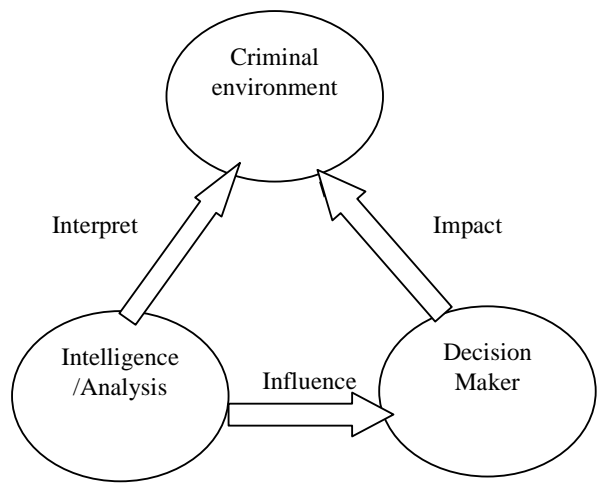

Figure 1: Pattern analysis theory

The most efficient and effective way of fighting crime today cannot be resourceful without geographical profiling. Criminal activities have become very complex in such a way that rapid monitory can only be achieved by using intelligent systems with geographical components."

Geographic profiling is a mathematical technique to derive information about a serial crime spree given the locations and times of Previous crimes in a given crime series [4]. Geographic profilers have access to a collection of strategies for predicting various attributes of crime such as a serial offender's home location, possible groups, relationship between events and crimes etc. These strategies range in complexity, some involve more calculations to implement than others and the assumption often made is that more complex strategies will outperform simpler strategies [5].

Over the years there have been developments and approaches in the analysis of crime data such as the introduction of a graph based dataset representation that allows one to mine a set of datasets for correlation [6], data mining techniques using clustering algorithm to help detect the crimes patterns and speed up the process of solving crime [7], procedure for detecting changes over time in the spatial pattern of point events, combining the nearest neighbor statistic and cumulative sum methods [8] etc.

Crime activities are geospatial phenomena and as such are geospatially, thematically and temporally correlated and. discovering these correlations allows a deeper insight into the complex nature of criminal behavior. This paper used Formal concept analysis, or Galois Lattices, a data analysis technique grounded on Lattice theory and propositional Calculus to discover the patterns and concepts within criminal data. This method considered the set of common and distinct attributes of crime data.

The organization of this paper is as follows, Section II proposes how the FCA was used to classify and analyze crime data. Section III provided application and results: provides the use of FCA crime analysis. The last section of this paper, section IV, Concludes the paper.

## II. METHODOLOGY

In this paper, Formal Concept Analysis, or Galois Lattices, a data analysis technique grounded on Lattice Theory and Propositional Calculus, was used for the visualization and analysis of crime data. This method considered the set of common and distinct attributes of crimes in such a way that categorization are done based on related crime types, geographical locations and the persons involved. Formal concept analysis (FCA) is a method of data analysis with growing popularity across various domains. FCA analyzes data and describes relationship between a particular set of objects and a particular set of attributes. Such data commonly appear in many areas of human activities. FCA produces two kinds of output from the input data. The first is a concept lattice. A concept lattice is a collection of formal concepts in the data which are hierarchically ordered by a subconcept-super concept relation [9][10].

In FCA, a formal context consists of a set of objects, $G$, a set of attributes, $M$, and a relation between $G$ and $M$, $I \subseteq G \times M$. A formal concept is a pair $(A,B)$ where $A \subseteq G$ and $B \subseteq M$. Every object in $A$ has every attribute in $B$. For every object in $G$ that is not in $A$, there is an attribute in $B$ that that object does not have. For every attribute in $M$ that is not in $B$ there is an object in $A$ that does not have that attribute. $A$ is called the extent of the concept and $B$ is called the intent of the concept.

If $g \in A$ and $m \in B$ then $(g,m) \in I$ ,or $gIm$.
A formal context is a tripel $(G,M,I)$, where
$G$ is a set of objects,
$M$ is a set of attributes
and $I$ is a relation between $G$ and $M$.
$(g,m) \in I$ is read as "object $g$ has attribute $m$".

For $A \subseteq G$, we define
$A´ := \{m \in M \mid \forall g \in A: (g,m) \in I \}$.
For $B \subseteq M$, we define dually
$B´ := \{g \in G \mid \forall m \in B: (g,m) \in I \}$.

For $A, A1, A2 \subseteq G$ holds:
$A1 \subseteq A2 \Rightarrow A`2 \subseteq A`1$
$A 1 \subseteq A``$
$A` = A```$

For $B, B1, B2 \subseteq M$ holds:
$B1 \subseteq B2 \Rightarrow B`2 \subseteq B`1$
$B \subseteq B``$
$B` = B```$

A formal concept is a pair $(A, B)$ where
$A$ is a set of objects (the extent of the concept),
$B$ is a set of attributes (the intent of the concept),
$A` = B$ and $B` = A$.

The concept lattice of a formal context $(G, M, I)$ is the set of all formal concepts of $(G, M, I)$, together with the partial order $(A1, B1) \leqslant (A2, B2): \Leftrightarrow A1 \subseteq A2 ( \Leftrightarrow B1 \supseteq B2)$.

The concept lattice is denoted by $\mathcal{B}(G,M,I)$.
*Theorem:* The concept lattice is a lattice, i.e. for two concepts $(A1, B1)$ and $(A2, B2)$, there is always
- greatest common subconcept: $(A1 \cap A2, (B1 \cup B2)´´)$
- and a least common superconcept: $((A1 \cup A2)´´, B1 \cap B2)$

More general, it is even a complete lattice, i.e. the greatest common subconcept and the least common superconcept exist for all (finite and infinite) sets of concepts.

Corollary: The set of all concept intents of a formal context is a closure system. The corresponding closure operator is $h(X) := X``$.

An implication $X \rightarrow Y$ holds in a context, if every object having all attributes in $X$ also has all attributes in $Y$.

Def.: Let $X \subseteq M$. The attributes in $X$ are independent, if there are no trivial dependencies between them.

## III. THE RESULTS AND ANALYSIS

Considering table 1 below, we have the rows to consist of persons and the columns which is the attributes, entails the age, sex, crime type committed by this persons, and the geographical location of the crime. Table 2 consists of the geographical locations and the economic factors, which includes income index, Index of education, and population index, existing at these locations. Figure two below is the geographical maps of the locations.

From Table1,
Let a = ages <18, b = ages <40 and c ages >40
For the sex, let m= male and f=female
Let c1=drugs, c2=rape, c3=burglary, c4=robbery
Let the geographical locations be denoted by g1, g2….g5.
From Table 2,
Let the geographical locations be denoted by g1, g2….g5.
For the index of income, let a be index =< 0.25, b be index = < 0.5, c be index =< 7.5, and d be index =< 1
For the index of education, let e be index =< 0.2, f be index = < 0.4, g be index =< 0.6, h be index =< 0.8, and i be index =< 1. Let that of population be {j, k, l, m, n}={0.2,0.4,0.6,0.8,1}.

TABLE 1: PERSONS X CRIME DATA WITH GEO. LOCATION

|   | Age | | | Sex | | Crime Type | | | | Geographical Location | | | | |
|---|---|---|---|---|---|---|---|---|---|---|---|---|---|---|
|   | a | b | c | m | f | c1 | c2 | c3 | c4 | g1 | g2 | g3 | g4 | g5 |
| P1 | x |   |   | x |   | x |   | x |   | x |   |   |   |   |
| P2 | x |   |   |   | x | x |   |   | x |   |   | x |   |   |
| P3 |   | x |   |   | x | x |   | x |   |   |   |   |   | x |
| P4 | x |   |   | x |   | x | x |   |   |   |   | x |   |   |
| P5 |   |   | x | x |   | x | x |   |   | x |   |   |   |   |
| P6 |   | x |   | x |   |   |   | x |   | x | x |   |   |   |
| P7 | x |   |   | x |   |   | x |   |   | x |   |   |   |   |
| P8 |   | x |   | x |   |   |   |   | x |   | x |   |   |   |
| P9 | x |   |   | x |   | x |   |   | x |   |   |   | x |   |

TABLE 2: GEOGRAPHICAL LOCATIONS X ECONOMIC FACTORS

|   | Income Index | | | | Index of Education | | | | | Population Index | | | |
|---|---|---|---|---|---|---|---|---|---|---|---|---|---|
|   | a | b | c | d | e | f | g | h | i | j | k | l | m | n |
| g1 | x |   |   |   | x |   |   |   |   |   | x |   |   |   |
| g2 |   | x |   |   | x |   |   |   |   |   |   | x |   |   |
| g3 | x |   |   |   | x |   |   |   |   |   |   |   | x |   |
| g4 | x |   |   |   | x |   |   |   |   |   |   |   |   | x |
| g5 |   |   | x |   | x |   |   |   |   | x |   |   |   |   |

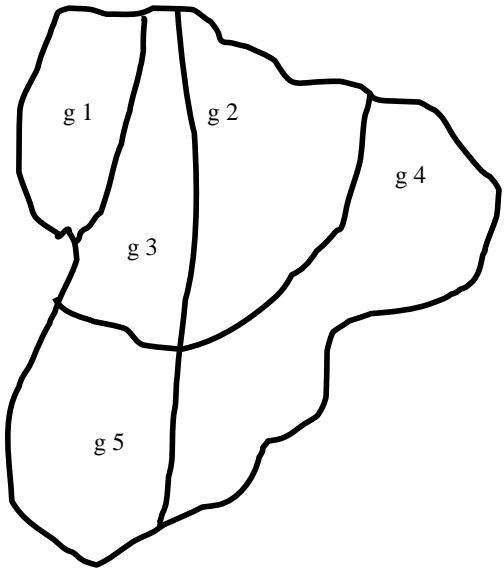

Figure 2: geographical map indicating the locations

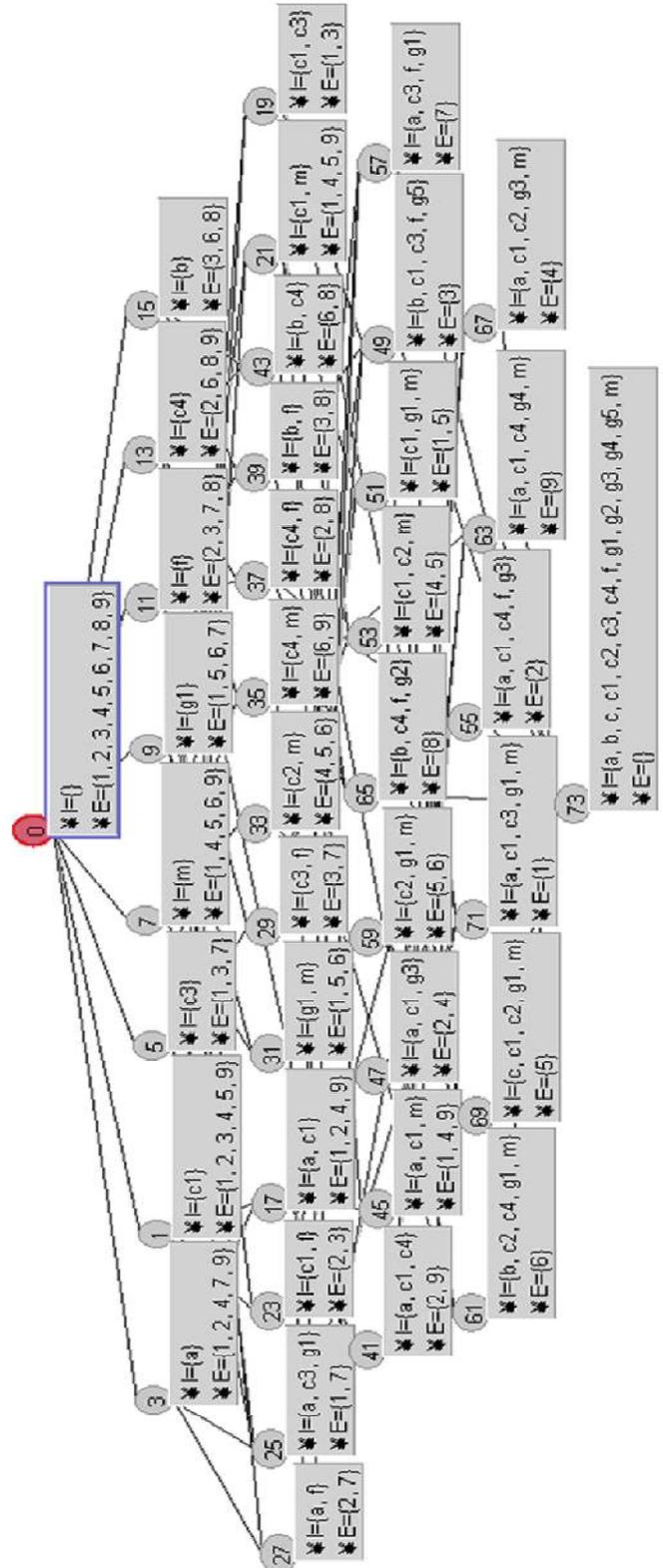

Figure 3: Galois lattices of intents and extents from table 1

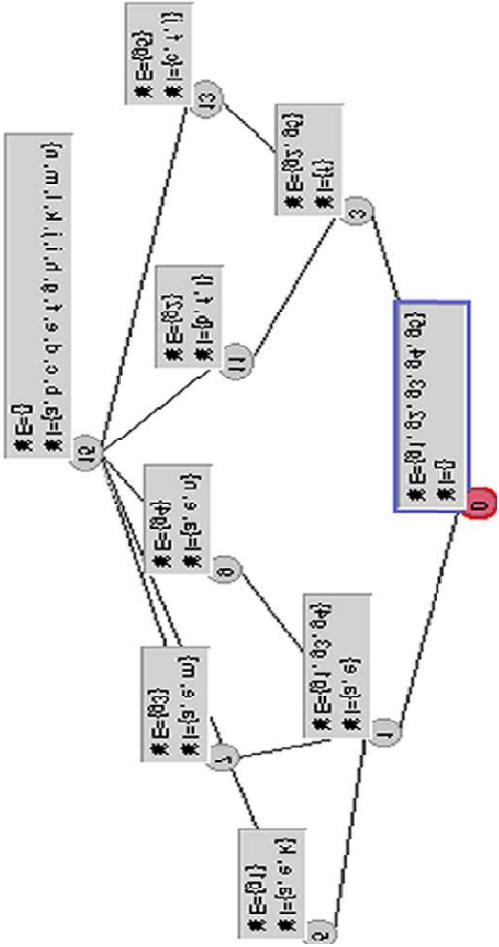

Figure 4: Galois lattices of intents and extents from table 2

TABLE 3: GEOGRAPHICAL LOCATIONS X CRIME TYPES

|    | c1 | c2 | c3 | c4 | Total |
|----|----|----|----|----|-------|
| g1 | 2  | 2  | 2  | 1  | 7     |
| g2 | 0  | 0  | 0  | 1  | 1     |
| g3 | 2  | 1  | 0  | 1  | 4     |
| g4 | 1  | 0  | 0  | 1  | 2     |
| g5 | 1  | 0  | 1  | 0  | 2     |
| Total | 6 | 3 | 3 | 4 | 16 |

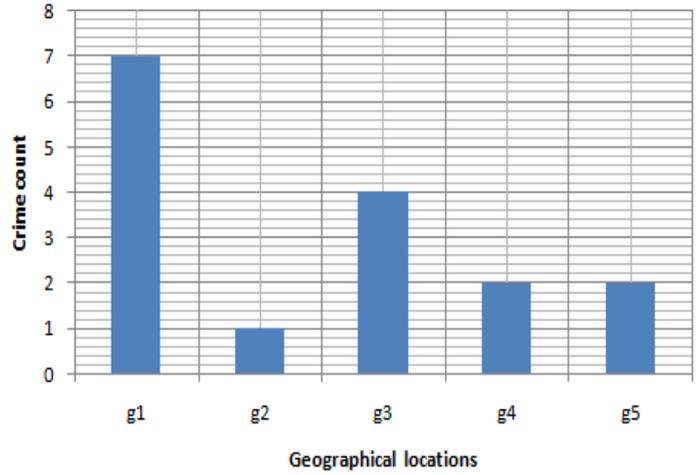

Figure 5: A graph of crime count against geographical locations.

The concept lattice of a formal context *(G, M, I)* as shown in figure 3, is the Galois lattices of intents and extents from table 1. This is the set of all formal concepts of *(G, M, I),* together with the partial order *(A1, B1)* ⩽ *(A2, B2):* ⇔ *A1 ⊆ A2 (⇔ B1 ⊇ B2).* From the concepts, it can be observed that concept 57, 61, 69, and 71 have most of the attribute g1 and most of the attributes of the crime types. This indicated the high occurrence of crime types in g1.

The concept lattice of a formal context *(G, M, I)* as shown in figure 4, is the Galois lattices of intents and extents from table 2. This is the set of all formal concepts of *(G, M, I),* together with the partial order *(A1, B1)* ⩽ *(A2, B2):* ⇔ *A1 ⊆ A2 (⇔ B1 ⊇ B2).* From the concepts, it can be observed that concept 1, and 5 have most of the attribute with the least index. This indicated the high occurrence of crime types in g1.

Table 3 consists of the matrix of geographical locations cross crime types and figure 5 is the graph of crime count versus geographical locations.

IV. CONCLUSION

A formal concept analysis was used to analyze and visualize crime data. And relationships between the various crime types and different geographical areas were visualized. This method considered the set of common and distinct attributes of crimes data in such a way that categorization was done based on related crime types and their geo-locations. This will help in building a more defined and conceptual systems for analysis of geographical crime data that can easily be visualized and intelligently analyzed by computer systems.